\documentclass[prd,11pt,tightenlines,nofootinbib,superscriptaddress]{revtex4}
\usepackage{amsfonts,amssymb,amsthm,bbm,amsmath,eufrak}
\usepackage{verbatim}
\usepackage{graphicx}
\usepackage{subfig}
\usepackage{tikz}

\usepackage{hyperref}
\hypersetup{
    colorlinks,
    citecolor=black,
    filecolor=black,
    linkcolor=black,
    urlcolor=black
}

\newcommand{\C}{{\mathbb C}}

\newcommand{\one}{\mathbbm{1}}

\newcommand{\cH}{{\mathcal H}}

\newcommand{\be}{\begin{equation}}
\newcommand{\ee}{\end{equation}}
\newcommand{\beq}{\begin{eqnarray}}
\newcommand{\eeq}{\end{eqnarray}}
\newcommand{\bea}{\begin{eqnarray}}
\newcommand{\eea}{\end{eqnarray}}

\newcommand{\bra}{\langle}
\newcommand{\ket}{\rangle}

\newcommand{\tr}{{\mathrm Tr}}

\newcommand{\rd}{\mathrm{d}}

\newcommand{\bpm}{\begin{pmatrix}}
\newcommand{\epm}{\end{pmatrix}}
\newcommand{\bvm}{\begin{vmatrix}}
\newcommand{\evm}{\end{vmatrix}}

\begin{document}

\title{A simpler way of imposing simplicity constraints}

\author{{\bf Andrzej Banburski}\email{abanburski@perimeterinstitute.ca}}
\affiliation{ Perimeter Institute for Theoretical Physics,
Waterloo, Ontario, Canada.}
\affiliation{Department of Physics, University of Waterloo, Waterloo, Ontario, Canada}

\author{{\bf Lin-Qing Chen}\email{lchen@perimeterinsititute.ca}}
\affiliation{ Perimeter Institute for Theoretical Physics,
Waterloo, Ontario, Canada.}
\affiliation{Department of Physics, University of Waterloo, Waterloo, Ontario, Canada}

\begin{abstract}
We investigate a way of imposing simplicity constraints in a holomorphic Spin Foam model that we recently introduced. Rather than imposing the constraints on the boundary spin network, as is usually done, one can impose the constraints directly on the Spin Foam propagator. We find that the two approaches have the same leading asymptotic behaviour, with differences appearing at higher order. This allows us to obtain a model that greatly simplifies calculations, but still has Regge Calculus as its semi-classical limit.
\end{abstract}

\maketitle

\section{Introduction}
One of the big open problems of modern theoretical physics is finding a non-perturbative definition of Quantum Gravity in 4 dimensions. Spin Foam models \cite{Rovelli:2011eq, Perez:2012wv} are an attempt at such a definition. The basic idea behind the framework is to start with a topological field theory, known as BF theory, which at the classical level can be constrained to a first-order formulation of General Relativity. More precisely, starting with the action
\be
S_{BF} = \int_\mathcal{M} B\wedge F
\ee
and imposing the simplicity constraint $B = \ast ( e\wedge e )+ \frac{1}{\gamma} e\wedge e$ reduces the theory to the Plebanski action for GR \cite{Plebanski:1977zz}. At the quantum level, the idea is to start with the well-known partition function of BF theory and impose the simplicity constraints on expectation values of states. In the spin representation of Spin Foam models, like the EPRL-FK models \cite{Freidel:2007py, Engle:2007uq, Engle:2007qf, Engle:2007wy}, the boundary states are given by spin networks, so the simplicity constraint is implemented on the boundary spin network of a fundamnetal building block -- a 4-simplex. 

Spin Foam models have been recently rewritten in a holomorphic representation using spinors  \cite{coh1, Livine:2007ya, Freidel:2009nu, Freidel:2009ck, Freidel:2010tt, Borja:2010rc, Livine:2011gp, Dupuis:2012vp, Dupuis:2010iq, Dupuis:2011dh, Dupuis:2011wy}. The standard framework for imposing the simplicity constraints was implemented in the holomorphic framework by Dupuis and Livine in \cite{Dupuis:2011fz}, which we will refer to as the DL model. In the Riemannian case of the gauge group $SU(2)_L\times SU(2)_R$, the constraint imposes left and right spinors to be proportional to each other on the boundary spinor network of a 4-simplex.

In a recent article \cite{Banburski:2014cwa} we have introduced a new Riemannian Spin Foam model in the holomorphic representation, which allowed the successful calculation of 4d Pachner moves. Rather than imposing the constraints on the boundary spinor networks, we impose them on the Spin Foam propagators. For the case of the 4-simplex, this effectively imposes the constraints not only on the boundary, but also in the bulk. This choice results in a reduction of two internal strands to one and allows to calculate the Spin Foam amplitudes much more efficiently.

Naively it is not obvious that imposing more constraints does not spoil the semi-classical limit of the Spin Foam model. In this article we calculate the asymptotic behaviour of the amplitude of a 4-simplex for both the  DL model  as well as the  new model proposed in \cite{Banburski:2014cwa}. We find first that the DL model's asymptotics turn out to be the same as the EPRL-FK model \cite{Bianchi:2008ae, Conrady:2008mk, Barrett:2009gg, Magliaro:2011dz, Han:2011rf,  Han:2011re, Han:2013gna, Han:2013hna, Han:2013ina, Hellmann:2013gva}, as was expected. Next, we study the asymptotics of the new model and find that due to non-trivial cancellations on-shell, it has the same 1st order behaviour, giving Regge Calculus \cite{Regge:1961px}. The differences in asymptotics between the models reside in the Hessian, the overall normalization and the higher order terms as well as on the off-shell trajectories. 
\section{Holomorphic Spin Foam Models}\label{sec:spinfoams}
In this section we will review the holomorphic Spin Foam models. We will start from a short review of the representation of SU(2) in terms of holomorphic functions of spinors, followed by a summary of the holomorphic simplicity constraints introduced by Dupuis and Livine in \cite{Dupuis:2011fz}. We will finish this section by showing how the constraints can be imposed in two ways -- on the boundary spinor network, as was done in the Dupuis-Livine model \cite{Dupuis:2011fz} and on the Spin(4) projectors, as we have introduced in \cite{Banburski:2014cwa}.

\subsection{Holomorphic representation}

Let us consider the Bargmann-Fock space \cite{Bargmann,Schwinger} of holomorphic functions on spinor space $\C^2$ endowed with the Hermitian inner product
\be \label{barg_in_prod}
  \bra f | g \ket = \int_{\C^2} \overline{f(z)} g(z) \rd\mu(z)
\ee
where $\rd\mu(z) = \pi^{-2} e^{-\bra z | z \ket}  \rd^{4}z$ and $\rd^{4}z$ is the Lebesgue measure on $\C^2$.
We use the notation
 $$|z\ket \equiv (\alpha, \beta )^t, \qquad |z] \equiv ( -\overline{\beta}, \overline{\alpha} )^t$$
 and $\check{z}$ to denote the conjugate spinor $|\check{z}\ket \equiv |z]$.
%
For the construction of Spin Foam amplitudes, we will be interested in the SU(2) invariant functions on $n$ spinors 
\be
f(g z_1, g z_2, ... , g z_n) = f(z_1, z_2, ... , z_n)\,, \quad \forall g \in {\rm SU(2)}.
\ee 
These invariant elements of $L^2(\C^2,\rd\mu)^{\otimes n}$ form a space $\cH_n = \bigoplus_{j_i} \cH_{j_1,...,j_n}$ of  $n$-valent intertwiners.
One way to construct an element of $\cH_n$ is to average a function of $n$ spinors over the group using the Haar measure.  In this way we can construct a projector $P :L^2(\C^2,\rd\mu)^{\otimes n} \rightarrow \cH_n$ as
\be
  P(f)(w_i) = \int \prod_i \rd\mu(z_i) P(\check{z}_i;w_i) f(z_1, z_2, ... , z_n) = \int_{\text{SU(2)}} \rd g f(g w_1, g w_2, ... , g w_n),
\ee

where the projection  kernel  in the $n=4$ case is  given by
\be
  P(z_i;w_i) = \int_{\text{SU}(2)} \rd g \,e^{\sum_i [z_i|g|w_i\ket} = 
\begin{tikzpicture}[baseline=0,scale=0.45]
  \node at (-4,1.5) {$[z_1|$}; \draw (-3,1.5) -- (-1,1.5);
  \node at (-4,0.5) {$[z_2|$}; \draw (-3,0.5) -- (-1,0.5);
  \node at (-4,-0.5) {$[z_3|$}; \draw (-3,-0.5) -- (-1,-0.5);
  \node at (-4,-1.5) {$[z_4|$}; \draw (-3,-1.5) -- (-1,-1.5);
  \draw (-1,-2) --(-1,2) -- (1,2) -- (1,-2) -- (-1,-2);
  \draw (1,1.5) -- (3,1.5); \node at (4,1.5) {$|w_1\ket$}; 
  \draw (1,0.5) -- (3,0.5); \node at (4,0.5) {$|w_2\ket$}; 
  \draw (1,-0.5) -- (3,-0.5); \node at (4,-0.5) {$|w_3\ket$}; 
  \draw (1,-1.5) -- (3,-1.5); \node at (4,-1.5) {$|w_4\ket$}; 
\end{tikzpicture}
\label{proj}
\ee
and $\rd g$ is the normalized Haar measure over SU(2).  In \cite{Freidel:2012ji} it was shown that the group integration can be performed explicitly and in \cite{Banburski:2014cwa} we have shown that the projector can rewritten as

\be \label{eqn_res_id_UN}
  P(z_i;w_i) = \sum_{J} \frac{\left( \sum_{i<j} [z_i|z_j\ket[w_i|w_j\ket\right)^J}{J!(J+1)!}.
\ee
We will also refer to the kernel $P(z_i;w_i)$ as a projector.
From the fact that $\int  \prod_i \rd\mu(w_i) P(z_i;w_i) P(\check{w}_i;z_i') =  P(z_i;z_i')$ we can construct coherent intertwiner states \cite{Freidel:2013fia}
\be\label{eq:cohstate}
  ( w_i \| j_i \, z_i \ket \equiv \int \rd g \prod_i \frac{[ w_i|g|z_i\ket^{2j_i}}{(2j_i)!}.
\ee
Spin Foam amplitudes are usually constructed as contractions of such coherent states. Before we construct these amplitudes, we have to discuss simplicity constraints.

\subsection{Simplicity Constraints}

In this section we will  review the idea behind simplicity constraints and their holomorphic version  \cite{Dupuis:2011dh}. The basic idea behind Spin Foam models is to take a topological field theory -- BF theory, and impose constraints to reduce it to General Relativity.

BF theory is defined on a principal bundle over a $d$ dimensional manifold $\mathcal{M}$, with a group $G$ and a connection $\omega$. B is a $d-2$ form in the adjoint representation of $G$. $F(\omega) = d \omega + \omega \wedge \omega$ is a curvature 2 form. The action is defined as 
\begin{equation}
S_{BF} = \int_{\mathcal{M}}tr\left( B\wedge F(\omega) \right)
\label{BF}
\end{equation}
It is a topological field theory in the sense that  all the solutions of equations of motion are locally gauge equivalent. One can prove  that a bivector $B^{IJ}$  in $\mathbb{R} ^4$ or $\mathbb{M}^{1,3}$ is a simple bivector if and only if there exists a vector $N^I$ such that $N_I B^{IJ} =0$. When this condition is satisfied,  $B^{IJ}$ is constrained to be  proportional to $e^I \wedge e^J$ or $\ast(e^I \wedge e^J) $. Using a parameter $\gamma$ (the Barbero-Immirzi parameter) to distinguish these two sectors, we obtain the Holst action for gravity
\begin{equation}
S_{Holst} = \int_{\mathcal{M}}tr\left( \ast(e \wedge e)\wedge F(\omega) + \frac{1}{\gamma} e \wedge e  \wedge F(\omega)\right).
\end{equation}

For the Riemannian 4d Spin Foam models, we use the gauge group $Spin(4) = SU(2)_L \times SU(2)_R$, which is the double cover of $SO(4)$. The holomorphic simplicity constraints are isomorphisms between the two representation spaces of $SU(2)$: for any two edges $i,j$ which connect to the same node $a$,
\begin{equation}
[z^a_{iL} | z^a_{jL} \ket = \rho^2 [z^a_{iR} | z^a_{jR} \ket
\label{hsc}
\end{equation}
where $\rho$ is related to the Barbero-Immirzi parameter by
\begin{equation}
\rho^2 = \left\{ 
  \begin{array}{ll}
 (1-\gamma)/(1+\gamma), & \quad  |\gamma| < 1\\
 (\gamma-1)/(1+\gamma), & \quad  |\gamma| > 1.
  \end{array} \right.
\end{equation}

The Eq.(\ref{hsc}) can be only satisfied if there exists a unique $\mathrm{SL}(2,\C)$  group element $g_a$ for each node $a$, such that
\begin{equation}\label{eq:simplicityonz}
\forall i, \ \ \  g_{a} |z_{iL}^a \ket = \rho\  |z_{iR}^a \ket.
\end{equation}
It is interesting to notice that $g_a$ can be  expressed purely in terms of left and right spinors as 
\begin{equation}
 g_{a} =\frac{ |z_{iR}^a \ket  \bra z^a_{iL}|  + | z^a_{iR} ] [z^a_{iL}|}{\sqrt{\bra z^a_{iL}|z^a_{iL} \ket  \bra z^a_{iR}|z^a_{iR} \ket }}, \ \ \ \forall i\in a
\label{g}
\end{equation}
The holomorphic simplicity constraints imply  the geometrical simplicity only when $g_a \in SU(2)$. This happens only when the closure constraints are satisfied, which in the holomorphic representation are imposed in the semi-classical limit, see  \cite{Dupuis:2011dh}.

Geometrically, each spinor defines a three vector $\vec V (z) \in \bold{R}^3$ through the equation,
\begin{equation}\label{eq:vecspin}
|z \ket \bra z| =\frac{1}{2} \left(  \one  \bra z | z \ket  + \vec V(z)  \cdot \vec\sigma \right),\ \ 
|z ] [ z| =\frac{1}{2} \left(\one   [ z | z ]   - \vec V(z)  \cdot \vec\sigma \right).
\end{equation}
 Thus around a node in a spin-network, each link (dual to a triangle in the simplicial manifold) is associated with two 3-vectors $\vec V_L(z)$ and $\vec V_R(z)$ given by the left and right spinors. These vectors correspond to the selfdual $b_+$ and anti-selfdual $b_-$ components of the $B$ field respectively :
\begin{equation}
V^i_L(z) = b_+^i := B^{0i} +\frac{1}{2} \epsilon^i_{kl} B^{kl}, \ \ \ \  V^i_R (z) =b_-^i :=  -B^{0i} +\frac{1}{2} \epsilon^i_{kl} B^{kl},
\end{equation}
At the level of the vectors  $\vec V_L(z)$ and $\vec V_R(z)$ the holomorphic simplicity constraints imply now
\begin{equation}
 g_{a} \triangleright  \vec V_L(z_{i}^a)  = \rho^2  \vec V_R(z_{i}^a),\ \ \ \ \ \forall i \in a
\end{equation}
which leads to the constraint that the norms of the selfdual and anti-selfdual components of the bivector $(g_{a}, \one) \triangleright  (B+ \gamma \ast B)\ $ have to be equal to each other:
\begin{equation}
|(1+\gamma) g_{a} \triangleright b_+| = |(1-\gamma)  \triangleright b_-|.
\end{equation}

Thus the $B$ field is a simple bivector, and for the node  $a$ there exists a common time norm to all the  bivectors:
\begin{equation}\label{eq:4dnormal}
\mathcal{N}_a = (g_{a}, \one)^{-1} \triangleright (1,0,0,0).
\end{equation}
This implies now that $B = \ast ( e\wedge e )+ \frac{1}{\gamma} e\wedge e$ should be satisfied in the semi-classical limit.

\subsection{Imposing constraints}
We will now impose the holomorphic simplicity constraints on the Spin(4) BF theory in order to obtain a model of 4d Euclidean quantum gravity. In the EPRL model the simplicity constraints
\be\label{eq:eprlconstraints}
j_l = \rho^2 j_r
\ee 
are imposed  on the intertwiners as an operator equation, providing a map from Spin(4) to SU(2), which can be graphically denoted as
\be
\raisebox{-10mm}{\includegraphics[keepaspectratio = true, scale =1.2] {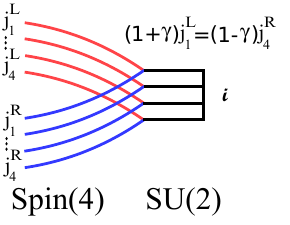}}
\ee
In the FK model \cite{Freidel:2007py}  it was shown however, that instead of working with the intertwiners, one can work with coherent states whose norm is an area of the faces of a tetrahedron. The Eq.(\ref{eq:eprlconstraints}) is satisfied by working with states $| j, \vec{n}; \rho^2 j, \vec{n}'\ket$ whose spin labels solve the constraints. Note however, that this approach to writing the coherent path integral doubles the number of variables, as one has to deal with independent vectors $\vec{n}$ and $\vec{n}'$. In \cite{Conrady:2009px} Conrady and Freidel extended the construction into fully geomterical coherent states by using the work of Guillemin and Sternberg \cite{GS}, which states that, for compact groups, ``quantization commutes with reduction". 

Here we will discuss two natural ways of imposing the holomorphic simplicity constraints. The first one is to impose the contraints on the boundary spinor network  defined by contraction of coherent states as done by Dupuis and Livine in \cite{Dupuis:2011fz}, to which we will refer to as the DL model. This corresponds to the following gluing of 4-simplices
\be\label{eq:dlconstraint}
\raisebox{-10mm}{\includegraphics[keepaspectratio = true, scale = 1.2] {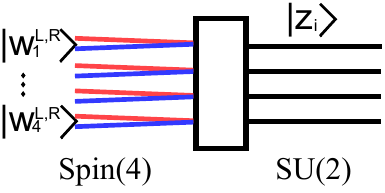}}
\ee
with two copies of spinors $|w^L\ket$ and $|w^R\ket$ on the inside of a 4-simplex and one copy satisfying $[z_L^i|z_L^j\ket = \rho^2 [z_R^i|z_R^j\ket$ on its boundary. In this way, DL model constitutes a weakening of the simplicity constraints, making all the constraints imposed coherently. This imposition of constraints is  dual to the FK one, as off-shell we have independent spins $j_L$ and $j_R$, but only one copy of spinors $|z\ket$. There are still however two independent copies of spinors $|w^L\ket$ and $|w^R\ket$.

The other way of imposing the holomorphic simplicity constraints inspired by the Guillemin-Sternberg result is to  impose the constraints on all the labels of the coherent states (\ref{eq:cohstate}), or effectively on the Spin(4) projector. This approach satisfactorily reduces the two copies of spinors to a single copy and gives the model we have recently introduced in \cite{Banburski:2014cwa}. Graphically this corresponds to
\be\label{eq:constrainnew}
\raisebox{-10mm}{\includegraphics[keepaspectratio = true, scale = 1.2] {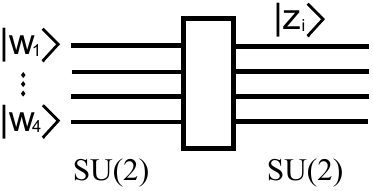}} 
\ee
with both $|z_L\ket = \rho |z_R\ket$ for the boundary spinors and $|w_L\ket = \rho |w_R\ket$ for the interior ones. In this way we obtain a model without the doubling of variables that the previous models exhibited, which in practice allows for much simpler calculations.

In this section we will give more details on these two alternatives before going on to show that indeed the two methods result in the same semi-classical limit.

\subsubsection{DL model  \label{sec:DLprojector}}

\begin{figure}[h]
	\centering
		\includegraphics[width=0.4\textwidth]{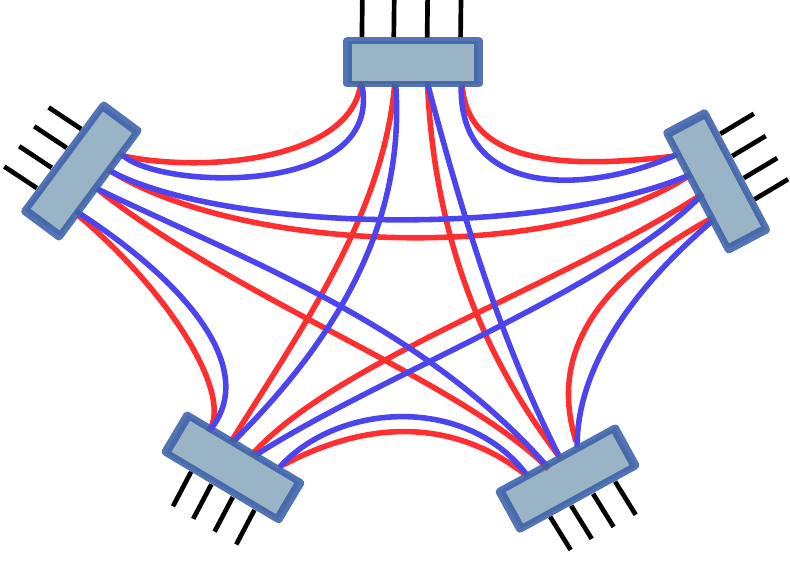}
	\caption{Graph for the 4-simplex amplitude in the DL model. The contractions inside correspond to two copies of BF 20j symbols, constrained on the boundary.}
		\label{fig:DLamplitude}
\end{figure}

In \cite{Dupuis:2011fz, Dupuis:2011dh} the simplicity constraints in the DL model are imposed on the boundary spinor network state, as is usually done in EPRL-FK models written in terms of coherent states. The amplitude for a single 4-simplex $\sigma$ is given by a product of contraction of coherent states for left and right sectors, with the simplicity constraints imposed on the boundary spinors as follows
\be\label{eq:dL4simplex}
\mathcal{A}_\sigma(\{z_\Delta^\tau\}) = \int \left[ \rd g^L_\tau\right]^5\left[ \rd g^R_\tau\right]^5 e^{\sum_{a,b\in\tau}\rho^2 [z_b^a|g_a^{L \, -1}  g_b^L|z_a^b\ket + [z_b^a|g_a^{R \, -1}  g_b^R|z_a^b\ket }
\ee
where $\tau$ is the set of tetrahedra labeled by $a, b$. To make the comparison of this imposition of constraints to the one we will introduce in the next section, we notice that the DL model can be rewritten as a contraction of a product of projectors $P_L( z_i, w_i^{L})P_R(\rho z_i, w_i^{R})$,
\be
P_{DL}(z_i;w^{L,R}_i)=\sum_{J_L, J_R} \frac{\left( \sum_{i<j} [z_i|z_j\ket[w^L_i|w^L_j\ket\right)^{J_L}}{J_L!(J_L+1)!} \frac{\left( \sum_{i<j} [z_i|z_j\ket[w^R_i|w^R_j\ket\right)^{J_R}}{J_R!(J_R+1)!} \rho^{2 J_R}
\ee
with $w^L$ and $w^R$ being two independent copies of spinors in the bulk, with no simplicity constraints imposed on them. This is exactly the expression we hinted to in  (\ref{eq:dlconstraint}) and it describes a mapping from the Spin(4) representation given by the spinors $|w^L\ket$ and $|w^R\ket$ into the SU(2) representation given by the spinor $|z\ket\equiv |z_L\ket$. The gluing of 4-simplex amplitudes in this language requires a product of two such projectors, giving a map $\textnormal{Spin(4)}\rightarrow\textnormal{SU(2)}\rightarrow\textnormal{Spin(4)}$, which can be graphically represented as

\be
\raisebox{-10mm}{\includegraphics[keepaspectratio = true, scale = 1] {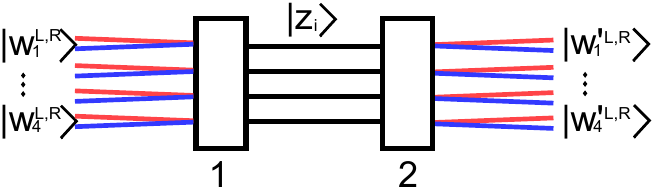}},
\ee
where the tetrahedra 1 and 2 belong to different 4-simplices. Using this graphical notation, the amplitude for a single 4-simplex then contains a double strands in the bulk corresponding to the two copies of spinors $|w^L\ket$ and $|w^R\ket$, and a single copy on the boundary corresponding to the spinors $|z\ket$. This can be seen in Fig. \ref{fig:DLamplitude}. 

\subsubsection{Constrained propagator}\label{sec:constrainedprojector}
Since spin foam amplitudes for BF theory are constructed from contractions of projectors  (\ref{proj}) into graphs corresponding to 4d quantum geometries, it is natural to impose the simplicity constraints directly onto the projectors themselves and hence on all the labels of the coherent states.  
Let us consider the Spin(4) projector obtained by taking a product of two SU(2) projectors
\be
  P(z^R_i;w^R_i) P(z^L_i;w^L_i) = \sum_{J_L} \frac{\left( \sum_{i<j} [z^L_i|z^L_j\ket[w^L_i|w^L_j\ket\right)^{J_L}}{J_L!(J_L+1)!} \sum_{J_R} \frac{\left( \sum_{i<j} [z^R_i|z^R_j\ket[w^R_i|w^R_j\ket\right)^{J_R}}{J_R!(J_R+1)!}
\ee
where we use a prime to distinguish the left and right 
SU(2) sectors.  We now impose the holomorphic simplicity constraints on both incoming and outgoing strands 
$$[z^L_i|z^L_j\ket = \rho^2 [z^R_i|z^R_j\ket \qquad [w^L_i|w^L_j\ket = \rho^2 [w^R_i|w^R_j\ket .$$ 
This makes the two products of spinors proportional to each other, with the proportionality constant being $\rho^4$, so we get that  the constrained projector is given by
\be\label{eq:projj}
  P_{\rho}(z_i;w_i)  = \sum_{J_L} \sum_{J_R} \frac{\rho^{4J_L}}{J_R!(J_R+1)!J_L!(J_L+1)!} \left( \sum_{i<j} [z_i|z_j\ket[w_i|w_j\ket\right)^{J_L+J_R},
\ee
where we have defined $|w\ket \equiv |w_R\ket$, $|z\ket\equiv |z_R\ket$. Note that by imposing the constraints, $P_\rho$ no longer satisfies the projection property $P \circ P = P$, hence we will refer to it as the propagator. This is the object we alluded to graphically in Eq. (\ref{eq:constrainnew}).
We can simplify this expression into a single sum by letting $J_L+J_R \rightarrow J$ to arrive at the most compact form of the constrained propagator
\be
  P_\rho(z_i;w_i)  = \sum_{J} F_\rho(J) \frac{\left( \sum_{i<j} [z_i|z_j\ket[w_i|w_j\ket\right)^{J}}{J!(J+1)!} \label{eq:PH}
\ee
where  we have recognized the power series expansion of the hypergeometric function
\be
 F_\rho(J)\equiv {}_2F_1(-J-1,-J;2;\rho^4) = \sum_{J'=0}^{J} \frac{\rho^{4J'}}{(J-J')!(J-J'+1)!J'!(J'+1)!}
\ee
We can now notice that the constrained Spin(4) propagator is just an SU(2) projector with non-trivial weights for each term that depend on the Barbero-Immirzi parameter. Note however that the power of $\rho$ in Eq.(\ref{eq:projj}) tracks the homogeneity of the left and right SU(2) sectors and as such preserves the Spin(4) invariance.

The imposition of simplicity constraints on all of the spinors has the additional effect of modifying the measure of integration on $\C^2$ 
\be
\rd\mu_\rho(z) := \frac{(1+\rho^2)^2}{\pi^{2}}e^{-(1+\rho^2)\bra z|z\ket} \rd^2 z
\ee

\begin{figure}[h]
	\centering
		\includegraphics[width=0.5\textwidth]{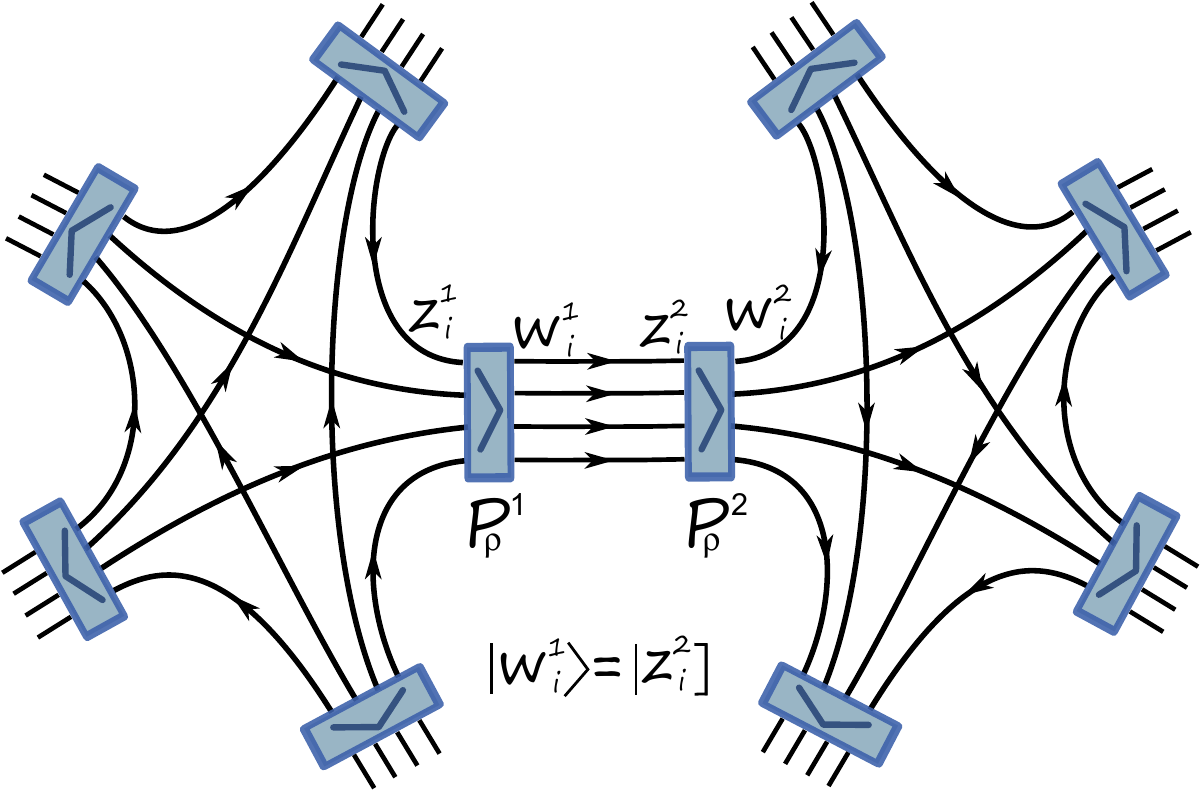}
	\caption{Graph for the amplitude of contraction of two 4-simplices. Propagators $P_\rho^1$ and $P_\rho^2$ belong to two different 4-simplices. The spinors on the same strand are contracted according to the orientation. For example, spinors $w^1_i = \check{z}^2_i$.}
		\label{fig:contraction}
\end{figure}

The partition function   can now be constructed from these constrained propagators. In \cite{Banburski:2014cwa} we proposed the amplitude
\be
\mathcal{Z}^{\Delta^\ast}_{G}=\sum_{j_f} \prod_{f \in \Delta^\ast}\mathcal{A}_f(j_f) \int \left\{\prod_{all}\rd\mu_\rho(z)\rd\mu_\rho(w)\right\}\sum_{k^e_{ff'}\in K_j}\prod_e  P_\rho^{k^e_{ij}}(z^e_i;w^e_i),
\ee
where $\mathcal{A}_f(j_f)$ is a face weight, the set $K_j$ is the set of integers $k_{ij}$ satisfying $\sum_{i\neq j}k_{ij}=2j_i$ and contraction of spinors according to the 2-complex $\Delta^\ast$ on different edges $e$ is implied. The $P_\rho^{k^e_{ij}}(z^e_i;w^e_i)$ is a constrained propagator at fixed spins and it is given by
\be
P_\rho^{k^e_{ij}}(z^e_i;w^e_i) :=\frac{ F_\rho(J_e)}{(J_e+1)!}\prod_{i<j}\frac{([z^e_i|z^e_j\ket[w^e_i|w^e_j\ket)^{k^e_{ij}}}{k^e_{ij}!} .
\ee
An example of an amplitude of two 4-simplices glued along one tetrahedron is shown in Fig. \ref{fig:contraction}. The main thing to note is that in this imposition of simplicity constraints, there is only one copy of strands, corresponding to spinors $|w\ket$, to be contracted in the bulk of the 4-simplex amplitude. The boundary data is the same as in the DL model and is given by the single copy of spinors $|z\ket$.

We thus see that the main difference between the two models is on the inside contraction in the 4-simplex. In the constrained propagator model, the simplicity constraints are imposed on both the boundary and the interior of the 4-simplex. In contrast, the standard approach is to have simplicity imposed only on the boundary, with the interior of the 4-simplex contractions identical to that of Spin(4) BF theory. The obvious worry one could have is that the constrained propagator model is over-constrained and does not lead to General Relativity in the semi-classical limit. We show in the next section however, that at least to the leading order, both models have the same asymptotic behaviour.

\section{Asymptotics}
In this section we will calculate the asymptotics of the two models with different imposition of simplicity constraints. First we show that the Dupuis-Livine model indeed has the same asymptotic behaviour as the EPRL-FK models. We then show that there are non-trivial cancellations in the asymptotic expansion of the constrained propagator model that lead to the same semi-classical limit as the DL model.

\subsubsection{The dihedral angle}
Before we calculate the asymptotic expansion of the Spin Foam amplitudes, we have to understand how to reconstruct from our data the angle appearing in the classical area-angle Regge action \cite{Dittrich:2008va}:
\be
S_{Regge} = \sum_{a<b} A_{ab} \xi_{ab} ,
\ee
where $A_{ab}$ is the area of face shared by tetrahedra $a$ and $b$, which share a common face with each other, and $\xi_{ab}$ is the 4-d dihedral angle, which is the angle between the two 4-vectors $\mathcal{N} _a, \mathcal{N} _b $ normal to the two tetrahedra $a, b$. 

We can find the expression for the 4-d dihedral angle using Eq. (\ref{eq:4dnormal}) from the section on simplicity constraints:
\begin{equation}
\begin{split}
\cos (\xi_{ab}) &=\mathcal{N} _a \cdot  \mathcal{N} _b\\
&=\frac{1}{2} tr\left[ g^{-1}_{a} \cdot g_{b} \right] =\frac{1}{2} tr\left[ g^{-1}_{b} \cdot g_{a} \right]
\end{split}
\end{equation}
Using the expression of eq.(\ref{g}), we can write the cosine of dihedral angle in terms of spinors,
\begin{equation}
\cos (\xi_{ab}) =\frac{[z^a_{iR}| z^b_{jR}\ket\bra z^b_{jL}|z^a_{iL}]  +\bra z_{iL}^a|z_{jL}^b\ket \bra z_{jR}^b|z_{iR}^a\ket  + c.c. }{2 \ |z^a_{iL}| |z^a_{iR}|  |z^b_{jL}| |z^b_{jR}| }
\label{cos}
\end{equation}
From the above two expressions, we can see that to decide the cosine of the dihedral angle $\xi_{ab}$, we need the data of two group elements  associated with two nodes (tetrahedra), or the data of both left and right spinors of any one strand from each of the two tetrahedra. In summary,
\begin{equation*}
\{g_a, g_b\} \rightarrow  \cos (\xi_{ab}), \  \text{or} \ \ \{z^a_{iR}, z^a_{iL}, z^b_{jR}, z^b_{jL}\} \rightarrow  \cos (\xi_{ab})  \ \ \forall i \in a, \forall j \in b
\end{equation*}

Let us recall additionally, that the models we consider have Spin(4) symmetry, so we can rotate these results by a Spin(4) transformation $G = (g_L, g_R)$. 
\subsubsection{The asymptotics of DL model}
An apparent difference between the holomorphic simplicity constraints and the ones in Euclidean EPRL/FK models is that they are constraints on spinors. However, they lead to the same constraint between spins,
\be
 \bra z_L | z_L\ket=  j_L = \rho^2 j_R  = \rho^2 \bra z_R | z_R\ket
\ee
for the coherent intertwiners in the large $|z|$ limit \cite{Dupuis:2011fz}.  In this section, we briefly show that for the amplitude of a 4-simplex, the DL model has the same action at critical points as EPRL/FK models for Barbero-Immirzi parameter $\gamma < 1$.

We can rewrite the amplitude (\ref{eq:dL4simplex}) of a 4-simplex $\sigma$ by expanding it in power series as
\begin{equation}\label{eq:dl4simpl}
\begin{split}
\mathcal{A}_\sigma &= \int \prod_{a<b} dg_{a,b}^{L,R} \  \bold{e}^{\rho^2 [z^a_{b}|g^{L \, -1}_a  g_b^L|z^b_{a}\ket +[z^a_{b}| g^{R \, -1}_a  g_b^R|z^b_{a}\ket}\\
&= \int \prod_{a<b} dg_{a,b}^{L,R} \sum_{j_{ab}^{L,R}} \frac{ \left( \rho^2 [z^a_{b}|g^{L \, -1}_a  g_b^L|z^b_{a}\ket\right)^{2j_{ab}^L} \left([z^a_{b}|  g^{R \, -1}_a  g_b^R|z^b_{a}\ket\right)^{2j_{ab}^R}}{(2j_{ab}^L)! (2 j_{ab}^R)!}.
\end{split}
\end{equation}
Now that we have made the summation over spins explicit, we can re-exponentiate this expression to  get the effective action of a 4-simplex amplitude $\mathcal{A}_\sigma =\sum_{j_{ab}^{L,R}}  \int \prod_a dg_{a}^{L,R}\  \bold{e}^{S_{eff}(j_{ab}^{L,R})}$ with
\begin{equation}
S_{eff}(j_{ab}^{L,R}) = \sum_{a,b \in \sigma} 2 j_{ab}^L \ln [z^a_{b}| g^{L \, -1}_a  g_b^L|z^b_{a}\ket+2 j_{ab}^R \ln [z^a_{b}| g^{R \, -1}_a  g_b^R|z^b_{a}\ket +N.
\end{equation}
where the numerical factor $N$ is given by
\begin{equation}
N=\sum_{a,b \in \sigma} 4j_{ab}^L \ln \rho - \ln(2j_{ab}^L)!- \ln(2j_{ab}^R)!
\end{equation}
It is important to note that this action is complex-valued. To study the asymptotic behaviour of the amplitude, we have to separate the real and imaginary parts. The real part of the action is
\begin{equation}
\begin{split}
\bold{Re} S_{eff}(j_{ab}^{L,R}) = &\sum_{a,b \in \sigma}  j_{ab}^L \ln \frac{1}{2} ( |z^a_{b}|^2 \ |z^b_{a}|^2 - (g_{a}^L \triangleright \vec V^a_{b} ) \cdot (g_{b}^L \triangleright \vec V^b_{a} ) )+ \\
&+ j_{ab}^R \ln \frac{1}{2} ( |z^a_{b}|^2 \ |z^b_{a}|^2 - (g_{a}^R \triangleright \vec V^a_{b} ) \cdot (g_{b}^R \triangleright \vec V^b_{a} ) )  +N.
\end{split}
\end{equation}
In the asymptotic analysis of complex functions the main contribution to the integral comes from critical points, which are stationary points of the action for which the real part is maximized. The critical point equations we get from variation of spinors $|z\ket$ are the closure constraints
\begin{equation}
\sum_{b\neq a} |z^a_{b} \ket \bra  z^a_{b} | = \sum_{b\neq a} j_{ab}^R \one
\label{closure}
\end{equation}
and  the orientation condition requiring certain vectors to be anti-parallel, which we get from the maximization of the real part of the action:
\begin{equation}
g_{a}^L \triangleright \hat v^a_{b} =- g_{b}^L \triangleright \hat v^b_{a}, \ \ g_{a}^R \triangleright \hat v^a_{b} =- g_{b}^R \triangleright \hat v^b_{a}, \ \ \text{where} \ \ \hat v = \vec{V}/|\vec{V}|.
\label{map}
\end{equation}
Using the relation (\ref{eq:vecspin}) between vectors and spinors, we find that these conditions imply  that the action of group elements on a spinor $z_a^b$ rotates it up to a phase into $\hat{z}_b^a$:
\begin{equation}
 g^{L \, -1}_a  g_b^L |z^b_{a}\ket=e^{i \phi ^{ab} _L}  |z^a_{b}],\ \ \  g^{R \, -1}_a  g_b^R |z^b_{a}\ket=e^{i \phi ^{ab} _R}  |z^a_{b}].
\label{phase}
\end{equation}
This implies that the following identity holds
\begin{equation}\label{eq:phiangles}
g^{R \, -1}_{b} g_{a}^R g^{L \, -1}_{a} g_{b}^L |z^b_{a}\ket=e^{i \phi ^{ab} _L-\phi ^{ab} _R} |z^b_{a}\ket .
\end{equation}

The reconstruction theorem from \cite{Barrett:2009gg} tells us now that given non-degenerate boundary data satisfying the closure constraint (\ref{closure}) and a set of group elements $g^{L,R}_a \in \textnormal{SU(2)}, a = 1,\ldots , 5$ solving the orientation condition (\ref{map}), we can reconstruct a geometric 4-simplex with the B field given by
\be\label{eq:reconstruction}
B_{ab} = \pm (j_{ab}^R + j_{ab}^L) (g^L_a, g^R_a)\triangleright ( v^a_{b}, v^a_{b}) ,
\ee
with the outward-pointing normal $\mathcal{N}_a$ obtained by acting with the Spin(4) element $ (g^L_a, g^R_a)$ on the vector $N_a = (1,0,0,0)$.

At this point, it is clear that the critical action of DL model is exactly the same as the one calculated in the asypmptotic analysis of the EPRL model in \cite{Barrett:2009gg}, and the imaginary part of the action reads 
\begin{equation}
\bold{Im} S_{eff}(j_{ab}^{L,R}) = \sum_{a,b \in \sigma} 2 j_{ab}^L \phi ^{ab} _L+2 j_{ab}^R \phi ^{ab} _R =\sum_{a,b \in \sigma}  k_{ab} ( \phi ^{ab} _L + \phi ^{ab} _R) + \gamma  k_{ab} ( \phi ^{ab} _R - \phi ^{ab} _L),
\end{equation}
where $k_{ab} = j_{ab}^L + j_{ab}^R$. To relate this to the area-angle Regge action, we have to relate the $\phi$'s to the dihedral angle. We cannot directly use our expression in Eq. (\ref{cos}) for the dihedral angle, since we no longer have the information about both the left and right spinors. We can however use the result of the reconstruction theorem from the Eq.(\ref{eq:reconstruction}) to construct the dihedral angle by the data $\{ g_{a}^R g^{L \, -1}_{a}, \ g_{b}^R g^{L \, -1}_{b}  \} $ as follows
\begin{equation}
\begin{split}
\cos (\xi_{ab}) &=\mathcal{N} _a \cdot  \mathcal{N} _b\\
& =\frac{1}{2} \tr\left[ g_{a}^R g_{a}^{L \, -1} \cdot g_{b}^L g_{b}^{R \, -1} \right]
\end{split}
\end{equation}
Notice however that we can obtain the same trace from the Eq. (\ref{eq:phiangles}), which tells us  that we can identify the cosine between the phase $(\phi ^{ab} _L-\phi ^{ab} _R)$ and the dihedral angle $\xi_{ab}$
\begin{equation}
\cos(\phi ^{ab} _L-\phi ^{ab} _R) = \cos(\xi_{ab}) .
\end{equation}
In \cite{Barrett:2009gg} it has been shown explicitly that the phase difference $(\phi ^{ab} _R-\phi ^{ab} _L)$ and the dihedral angle $\xi_{ab}$ can be identified up to a $\pm$ sign, which is due to the relative orientation of the bivector and 4-simplex. The  angle $(\phi ^{ab} _L + \phi ^{ab} _R)$ can be shown to be proportional to $2 \pi$ \cite{Barrett:2009gg}.

Hence the semi-classical limit of the Dupuis-Livine model is the same as the EPRL-FK models and is given by the action
\be
S = \sum_{a,b \in \sigma}\gamma  k_{ab} \xi_{ab} .
\ee
Since in Loop Quantum Gravity the spectrum of the area operator is given by $A_j = \gamma \sqrt{j(j+1)}$, in the large spin limit  we have obtained exactly the area-angle Regge action \cite{Regge:1961px, Dittrich:2008va}.

\subsubsection{The asymptotics of constrained propagator model}
Let us now finally show that the constrained propagator model also leads to the same semi-classical limit as the EPRL-FK models. We first have to rewrite the amplitude in terms of group variables. Recall that we can write an SU(2) propagator as
\be
  P(z_i;w_i) = \int_{\text{SU}(2)} \rd g \,e^{\sum_i [z_i|g|w_i\ket}
\ee
Thus taking two copies of such projectors and constraining them both in the $|w\ket$ and in the $|z\ket$ spinors, we get that the constrained propagator (\ref{eq:PH}) can be written as
\be\label{eq:constwithg}
P_\rho (z_i;w_i) = \int_{\text{SU}(2)_L\times \text{SU}(2)_R}\rd g^L \rd g^R \bold{e}^{\sum_i [z_i|g^R + \rho^2 g^L|w_i\ket}.
\ee
The 4-simplex amplitude is now just a simple contraction of 5 such propagators. To compare it however to the amplitude in the DL model, we have to perform the integrate out the $|w_i\ket$ spinors in order to have the same number of variables. After the $|w_i\ket$ integration, the amplitude becomes
\be
\tilde{\mathcal{A}}_\sigma = \int \prod_a dg_{a}^{L,R} \  \bold{e}^{(1+\rho^2)^{-1}[z^a_{b}| (  g^{R \, -1}_{a}+ \rho^2   g^{L \, -1}_{a} )( g_{b}^R+\rho^2   g_{b}^L )|z^b_{a}\ket}.
\ee
We can see that there is a mixing between left and right sectors -- while in the DL model the left and right group elements $g^L, \ g^R$ are multiplied separately as in Eq.(\ref{eq:dl4simpl}), here the relevant group elements become a combination $(g^R + \rho^2 g^L)$. Expanding this in a power series it would seem we would get four independent terms. However, since in the large $z$ limit the holomorphic simplicity constraints imply that we have $j^L=\rho^2 j^R$, one can show that only three summations are independent, so the amplitude can be written as
\begin{equation}\label{eq:consamp}
\begin{split}
\tilde{\mathcal{A}}_\sigma =& \int \prod_a dg_{a}^{L,R} \sum_{j_{ab}^{L,R}, J_{ab}} \frac{([z^a_{b}|  g^{R \, -1}_{a} g_{b}^R |z^b_{a}\ket)^{2j_{ab}^R - 2J_{ab}}}{(2j_{ab}^R - 2J_{ab})!}  \frac{(\rho^4 [z^a_{b}|     g^{L\, -1}_{a} g_{b}^L |z^b_{a}\ket)^{2j_{ab}^L - 2J_{ab}}}{(2j_{ab}^L - 2J_{ab})!} \times\\
& \times\frac{(\rho^2 [z^a_{b}|  g^{R\, -1}_{a} g_{b}^L |z^b_{a}\ket)^{2J_{ab}}}{(2J_{ab})!}
 \frac{(\rho^2 [z^a_{b}| g^{L\, -1}_{a} g_{b}^R |z^b_{a}\ket)^{2J_{ab}}}{(2J_{ab})!} (1+\rho^2)^{-2(j_{ab}^L+j_{ab}^R)},
\end{split}
\end{equation}
with the the spins satisfying
\be
j_{ab}^R \geq J_{ab}, \ \ \ \ \ j^L_{ab} \geq J_{ab}.
\ee
This means that the mixed left-right terms never overtake the pure left and right sectors. For the  details of this calculation, see the Appendix.

We thus get that the effective action of the constrained propagator model for a single 4-simplex is simply
\begin{equation}
\begin{split}
\tilde{S}_{eff}(j_{ab}^{L,R}, J_{ab})& = \sum_{a,b \in \sigma} 2 (j_{ab}^R -J_{ab}) \ln [z^a_{b}| g^{R\, -1}_{a} g_{b}^R|z^b_{a}\ket+2( j_{ab}^L-J_{ab}) \ln [z^a_{b}| g^{L\, -1}_{a} g_{b}^L|z^b_{a}\ket \\
& \underbrace{+2J_{ab} \ln [z^a_{b}| g^{R\, -1}_{a} g_{b}^L |z^b_{a}\ket+2J_{ab} \ln [z^a_{b}| g^{L \, -1}_{a} g_{b}^R  |z^b_{a}\ket}_{\text{mixed}}
+ \tilde{N},
\end{split}
\end{equation}
where the numerical factor $\tilde{N}$ carries all the normalization factors and is a function of the different spins and $\rho$ given by
\begin{equation}
\tilde{N}=\sum_{a,b \in \sigma} 8 j_{ab}^L \ln \rho - 2 (j_{ab}^L+j_{ab}^R)\ln(1+\rho^2) - \ln(2j_{ab}^L-2J_{ab})!- \ln(2j_{ab}^R-2J_{ab})!-2\ln(2J_{ab})!
\end{equation}
We can see that compared with the DL model, the effective action of the constrained propagator model has two additional terms which are underbraced and an additional spin $J_{ab}$. Nonetheless, we again obtain the closure equation  from the variation of spinor $|z\ket$,
\be
\sum_{b\neq a} |z^a_{b} \ket \bra  z^a_{b} | = \sum_{b\neq a} j_{ab}^R \one .
\ee
To see how these additional terms change the asymptotics, let us examine the terms in the real part of this action
\begin{equation}
\begin{split}
&( j_{ab}^L - J_{ab}) \ln \frac{1}{2} \left( |z^a_{b}|^2  |z^b_{a}|^2 - (g_{a}^L \triangleright \vec V^a_{b} )\! \cdot\! (g_{b}^L \triangleright \vec V^b_{a} ) \right)+ 
 (j_{ab}^R-J_{ab}) \ln \frac{1}{2}  \left( |z^a_{b}|^2  |z^b_{a}|^2 - (g_{a}^R \triangleright \vec V^a_{b} ) \!\cdot\! (g_{b}^R \triangleright \vec V^b_{a} ) \right) \\
& + J_{ab} \ln\frac{1}{2} \left( |z^a_{b}|^2 \ |z^b_{a}|^2 - (g_{a}^R \triangleright \vec V^a_{b} ) \cdot (g_{b}^L \triangleright \vec V^b_{a} ) \right) + J_{ab} \ln \frac{1}{2} \left( |z^a_{b}|^2 \ |z^b_{a}|^2 - (g_{a}^L \triangleright \vec V^a_{b} ) \cdot (g_{b}^R \triangleright \vec V^b_{a} ) \right)  + \tilde{N}.
\end{split}
\end{equation}
 At the critical points, we also require the real part of the effective action to be maximized. Since the real part of the action can be written as $\bold{Re} \tilde{S}_{eff} = S_{LL}+S_{RR}+S_{RL}+S_{LR}$ and all the coefficients in front of the logarithms are positive,  the maximization condition implies that all the four terms have to be maximized independently.  Thus the following critical equations substitute the Eq.(\ref{map}) in DL model,
\begin{equation}
g_{a}^L \triangleright \hat v^a_{b} =- g_{b}^L \triangleright \hat v^b_{a} = g_{a}^R \triangleright \hat v^a_{b} =- g_{b}^R \triangleright \hat v^b_{a}, \ \ \text{where} \ \ \hat v = \vec{V}/|\vec{V}|.
\end{equation}
When written in terms of spinors $|z\ket$ and $|z]$, this means that apart from the spinorial orientation condition in Eq.(\ref{phase}),
\begin{equation*}
g^{L\, -1}_{a}g_{b}^L |z^b_{a}\ket=e^{i \phi ^{ab} _L}  |z^a_{b}],\ \ \ g^{R\, -1}_{a}g_{b}^R |z^b_{a}\ket=e^{i \phi ^{ab} _R}  |z^a_{b}],
\end{equation*}
 relating  $|z_a^b\ket$ to  $|z_b^a]$ up to a phase, we also have two additional phases $\psi$ and $\theta$ appearing between the mixed left-right terms
\begin{equation}
g^{L\, -1}_{a}g_{b}^R |z^b_{a}\ket=e^{i \psi ^{ab}}  |z^a_{b}],\ \ \ g^{R\, -1}_{a}g_{b}^L |z^b_{a}\ket=e^{i \theta^{ab}} |z^a_{b}].
\label{phase2}
\end{equation}
Let us now plug in the critical point equations (\ref{phase}) and (\ref{phase2}) into the the effective action to find the semi-classical behaviour of the amplitude. The  imaginary part of the effective action becomes a function of three spins and four angles, given by
\begin{equation}
\bold{Im} \tilde{S}_{eff}(j_{ab}^{L,R}, J_{ab}) = \sum_{a,b \in \sigma} 2 (j_{ab}^L-J_{ab}) \phi ^{ab} _L+2 (j_{ab}^R-J_{ab}) \phi ^{ab} _R  +2J_{ab} (\psi ^{ab} +\theta^{ab})
\end{equation}
At first sight this is quite different from the effective action of the DL model, with two extra angles and an additional spin label to sum over. Let us notice however, that using the critical point equations (\ref{phase}) and (\ref{phase2}), we can get the relation
\begin{equation}
g^{L\, -1}_{b}g_{b}^R |z^b_{a}\ket = e^{i (\psi ^{ab} - \phi ^{ab}_L)}  |z^a_{b}] = e^{i (\phi ^{ab}_R - \theta^{ab}) }  |z^a_{b}] .
\end{equation}
This condition implies that the additional angles $\psi$ and $\theta$ we had to introduce are actually related to the angles $\phi_L$ and $\phi_R$ by
\begin{equation}
\psi ^{ab}  + \theta^{ab} = \phi ^{ab}_L +\phi ^{ab}_R  \ \ \text{mod} \ 2\pi .
\end{equation}
This is exactly the combination of angles that allows us to drop the terms proportional to $J_{ab}$ in the action. Hence we have that the imaginary part of the effective action is exactly the same as the one in DL model,
\begin{equation}
\bold{Im} \tilde{S}_{eff}(j_{ab}^{L,R},J_{ab}) = \sum_{a,b \in \sigma} 2 j_{ab}^L \phi ^{ab} _L+2 j_{ab}^R \phi ^{ab} _R=\bold{Im} S_{eff}(j_{ab}^{L,R}) 
\end{equation} 
The rest of the asymptotic analysis of this action carries over in exactly the same way, as in the EPRL-FK models. Thus we have proved that the constrained propagator model has in the asymptotic expansion the same effective action as the DL model, which in turn has the same semi-classical limit as the EPRL-FK models. 

It is important to note here that in the case of both of the models we have not performed the full asymptotic analysis, which would require the calculation of the Hessian, as it is not necessary for establishing that the models are described by Regge Calculus in the semi-classical  limit. We expect that  where the two models show differences is exactly in the Hessian and the overall normalization as well as possibly in the higher order terms in the asymptotic expansion. 

\section{Conclusions}
In this article we have studied the asymptotic expansion of 4-simplex amplitudes for the DL Spin Foam model and the newly introduced constrained propagator model. In the large $|z|$ limit (corresponding to the usual large-j limit in the spin representation) we have found that the DL model has the same first order expansion as the EPRL-FK Riemannian model. We have also shown that the constrained propagator model has a different amplitude for a 4-simplex, which however agrees with the DL model's one on-shell. We expect the differences to show up in the Hessian matrix and in the higher order terms. Hence both models lead to Regge Calculus in the semi-classical limit. 

The results obtained here prove the statements made in \cite{Banburski:2014cwa} and justify the use of the constrained propagator model. Compared to the DL model, the calculations are simplified greatly when done in terms of spinors (with the group elements integrated out), due to the smaller number of internal strands. Since the models based on the linearized simplicity constraints have the same leading semi-classical behaviour, one needs a different criterion for choosing which model to work with. An obvious one is simplicity and usability -- using the technique of the homogeneity map introduced in \cite{Banburski:2014cwa}, the calculations of Spin Foam amplitudes are finally tractable. Another criterion could be the divergence properties of the model. Naively, one could expect the more constrained model to be less divergent. Using the homogeneity map actually allows us to calculate exactly the divergence of an arbitrary Spin Foam amplitude in the constrained propagator model \cite{Chen:2015}.

An intriguing question is the extension of these results to the Lorentzian case. The big challenge to a trivial extension is the fact, that there is no known unitary holomorphic representation of the $SL(2,\C)$ group. This means that there is no obvious way of rewriting integrals of Wigner matrices into Gaussians. However, one could explore the possibility of adapting the homogeneity map to some other functions  that have the orthogonality property. Moreover, the holomorphic simplicity constraints have been rewritten in a similar way for the Lorentzian case \cite{Dupuis:2011wy}, using twistors instead of spinors, so defining a constrained propagator model for  $SL(2,\C)$ seems to be a possibility. We leave this construction to future research.

\acknowledgments

We would like to thank Laurent Freidel and Jeff Hnybida for comments and discussions, as well as for collaboration on the bigger project this article is a part of. Research at Perimeter Institute is supported by the Government of Canada through Industry Canada and by the Province of Ontario through the Ministry of Research and Innovation.

\appendix

\section{The 4-simplex amplitude for the constrained propagator model}
This appendix shows that in the constrained propagator model, the amplitude for a 4-simplex can be indeed written as (\ref{eq:consamp}). Starting from the constrained propagators (\ref{eq:constwithg}), we can straighforwardly  get that the spin foam amplitude for a single 4-simplex in the constrained propagator model is given by
\begin{equation}
\tilde{\mathcal{A}}_\sigma = \int \prod_a  d\mu_\rho(w^a_{b}) dg_{a}^{L,R} \ \ \bold{e}^{\tau^a_{bR} [w^a_{b}| g_{a}^R| z^a_{b}\ket + \tau^b_{aR} \bra w^a_{b}| g_{b}^R| z^b_{a}\ket +\tau^a_{bL} \rho^2 [w^a_{b}| g_{a}^L| z^a_{b}\ket+\tau^b_{aL} \rho^2 \bra w^a_{b}| g_{b}^L| z^b_{a}\ket }
\end{equation}
where we have added a collection of $\tau$'s which all have trivial values 1. However,  under Taylor expansion, the power of $\tau$ gives the homogeneity of the corresponding term. For example, the first term in the exponent becomes
\be
\sum_{j_a^R}\frac{(\tau^a_{bR}[w^a_{b}| g_{a}^R| z^a_{b}\ket ) ^{2 j_{a}^R}}{(2 j_{a}^R)!}.
\ee
We get similar expressions for all the different $\tau$'s, which are raised to the appropriate powers in their series expansions,
\begin{equation}\label{eq:taus}
(\tau^a_{bR}) ^{2 j_{a}^R}, \ (\tau^a_{bL} )^{2 j_{a}^L}, \ (\tau^b_{aL} )^{2 j_{b}^L}, \ (\tau^b_{aR}) ^{2 j_{b}^R}.
\end{equation}
The reason for temporarily introducing these factors of $\tau$ is  that before we continue with the asymptotics, we have to reduce the action to the same number of variables, as in the DL model's one.  To do this we have to integrate out the auxilliary contracting spinors $w_a^b$. We will see however, that this produces a mixing between the different  spins, so the factors of $\tau$ keep track of the spin information. After the $|w\ket$ integration, the amplitude becomes
\be
\tilde{\mathcal{A}}_\sigma = \int \prod_a dg_{a}^{L,R} \  \bold{e}^{(1+\rho^2)^{-1}[z^a_{b}| ( \tau^a_{bR} g^{R \, -1}_{a}+ \rho^2 \tau^a_{bL}  g^{L \, -1}_{a} )(\tau^b_{aR} g_{b}^R+\rho^2 \tau^b_{aL}  g_{b}^L )|z^b_{a}\ket}.
\ee
 Expanding this expression in a power series we get again four terms, but with different expansion coefficients:
\begin{equation}\label{eq:amplitudetaus}
\begin{split}
\tilde{\mathcal{A}}_\sigma &= \int \prod_a dg_{a}^{L,R} \sum_{j_i} ([z^a_{b}| \tau^a_{bR} \tau^b_{aR} g^{R \, -1}_{a} g_{b}^R |z^b_{a}\ket ^{2j_1}  (\rho^4 [z^a_{b}|  \tau^a_{bL} \tau^b_{aL}   g^{L\, -1}_{a} g_{b}^L |z^b_{a}\ket)^{2j_2} \times\\
& \times(\rho^2 [z^a_{b}| \tau^a_{bR}\tau^b_{aL} g^{R\, -1}_{a} g_{b}^L |z^b_{a}\ket)^{2j_3}
 (\rho^2 [z^a_{b}| \tau^b_{aR}\tau^a_{bL} g^{L\, -1}_{a} g_{b}^R |z^b_{a}\ket)^{2j_4})/ \prod_i (2j_i)!(1+\rho^2)^{2 j_i}.
\end{split}
\end{equation}
By comparing the coefficients of the power expansion of the different $\tau$'s in Eq. (\ref{eq:taus}) and (\ref{eq:amplitudetaus}), we find that $j_i$ and $j_{a,b;L,R}$ are related by the following set of equations:
\begin{equation}
 \left\{ 
  \begin{array}{llll}
 j_1+j_3 = j_{a}^R\\
 j_1+j_4 = j_{b}^R\\
j_2+j_4=j_{a}^L\\
j_2+j_3 =j_{b}^L.
  \end{array} \right.
\end{equation}
Note now that in the large $z$ limit the holomorphic simplicity constraints imply that we have $(j_{a}^L,j_{b}^L )=\rho^2 (j_{a}^R,j_{b}^R)$, which allows us to eliminate one of the spins and leads to an important relation for the asymptotic analysis:
\begin{equation}
j_3=j_4,\ \ j_{a}^R=j_{b}^R:=j_{ab}^R,\ \  j_{a}^L=j_{b}^L:=j_{ab}^L
\end{equation}
In the main text we define $J_{ab} \equiv j_3$. After the $\tau$'s have completed their mission, we can discard them by setting their values back to 1. Thus we have proved that the amplitude is indeed (\ref{eq:consamp}).


\end{document}